\documentstyle[12pt,psfig]{article}
\newcommand{\onlinecite}{\cite}
\begin{document}
\title{\Large{\bf{
      \quad\\
      The Equilibrium Shape of Quantum Dots
      }}}
\author{ \normalsize
	 \quad\\
         \normalsize 
	 E. Pehlke,\thanks{Present address: Physik-Department T30,
                   Technische Universit\"at M\"unchen, D-85747 Garching,
                   Germany.}\quad
         N. Moll,\thanks{Present address: Department of Physics,
                  Massachusetts Institute of Technology,
                  Cambridge, MA 02139, USA.}\quad
         and M. Scheffler\\
	 \quad\\ \quad\\ 
	 \normalsize
         Fritz-Haber-Institut der Max-Planck-Gesellschaft,\\
	 \normalsize
         Faradayweg 4--6,
         D-14195 Berlin-Dahlem, Germany
      }
\date{}
\maketitle
\centerline{\bf Abstract} 
The formation of dislocation-free three-dimensional islands
during the heteroepitaxial growth of lattice-mismatched materials
has been observed experimentally for several material systems. 
The equilibrium shape of the islands
is governed by the competition between the surface energy and the elastic
relaxation energy of the islands as compared to the uniform strained
film.
As an exemplification we consider the experimentally intensively investigated 
growth of InAs quantum dots
on a GaAs(001) substrate, deriving the equilibrium shape as a function of
island volume. For this purpose InAs surface energies have been 
calculated within
density-functional theory, and a continuum approach has been applied to 
compute the elastic relaxation energies.

\section*{\normalsize \bf 1. Introduction}

When a material\footnote{
	Here we consider the situation that the surface energy of the 
	growing material is smaller than the surface energy of the substrate, 
	resulting in the formation of a wetting layer and the Stranski-Krastanov
	growth mode.} 
is grown on a lattice-mismatched substrate, the uniform 
strained film becomes unstable beyond some critical thickness.
Strain relaxation can be achieved by the introduction of dislocations.
However, there is another important and general mechanism of strain relaxation. 
For Ge/Si(001) Eaglesham and Cerullo\cite{eaglesham:90} observed 
the formation of three-dimensional Ge islands that are
dislocation-free. This change in surface morphology is driven by
the gain in elastic relaxation energy of the islands, which overcompensates
the energetical cost due to the increase of surface area. 
Provided the interface energy for dislocated islands is sufficiently large
this was shown 
to yield the energetically preferred morphology as opposed to
both the uniform film and dislocated islands for small 
island size.\cite{vanderbilt:91}

Formerly, the roughening of the surface during Stranski-Krastanov growth
was considered a nuisance for device fabrication. Nowadays, the
strain-induced self-assembly of small three-dimensional islands 
is in fact exploited 
to produce ordered arrays of quantum dots.\cite{ledentsov:96}
A frequently studied 
example\cite{snyder:91,leonard:93,moison:94,ruvimov:95,bressler:95,petroff:96} 
is InAs/GaAs(001), the lattice mismatch
amounting to about 7\%.
Quantum dots have attracted great interest due to their zero-dimensional
electronic density of states.\cite{grundmann:95a,grundmann:95b} 
They may even be of tech\-no\-lo\-gical importance; e.g., 
buried quantum dots are advantageous for
improving the device properties of semiconductor lasers.\cite{kirstaedter:94}

The dispersion of the sizes of the InAs/GaAs(001) quantum dots 
is remarkably low.\cite{moison:94}
This represents an important and desirable feature of quantum dot growth
both in view of measurements and device applications, 
because the delta function shaped density 
of states of every single quantum dot is 
smeared out by ensemble averaging over the size distribution.
However, despite its potential importance for the optimization of the quality
of the quantum dot arrays, the details of the growth mechanism 
are not yet well understood. In particular, the reason behind the 
narrow size distribution, i.e. even the question whether it is due 
to kinetics or energetics, is still 
controversial.\cite{priester:95,shchukin:95a,jesson:96a} 

In this article we will focus on the equilibrium shape of 
coherent three-dimensional InAs islands. Knowledge of the 
equilibrium shape as a function of volume is
an essential prerequisite for deciding upon the true growth mechanism. 
Though kinetic effects will be important, we expect 
three-dimensional islands in equilibrium to be observable under
appropriate experimental conditions: When the concentration of quantum
dots is low, diffusion of atoms on a single island should be faster 
than material exchange between the islands. Thus shape equilibration
should occur on a timescale faster than that of Ostwald ripening.
Conversely, if already the experimentally observed shape deviates from
the equilibrium shape, equilibrium thermodynamics will not be adequate to
describe the island size distribution under the respective 
growth conditions.

To calculate the equilibrium shape we have computed InAs surface energies
for several surface orientations using density-functional theory.
The elastic relaxation energy is calculated within continuum theory,
applying a finite element approach. The total energy of the islands,
including the strain field in the substrate, is given by the sum
of these two energy contributions. 
All the more delicate effects,\cite{shchukin:95a} like 
the strain-dependence of the surface energy and island-island interaction
are neglected for the purpose of predicting island shapes at fixed volume.

\section*{\normalsize \bf 2. Surface Energy }

To determine the InAs surface energies 
we have carried out total-energy calculations within density-functional 
theory. The local-density approximation is applied to the
exchange-correlation energy-functional, using Perdew and Zunger's\cite{perdew:81}
parameterization of Ceperley and Alder's\cite{ceperley:80} data 
for the correlation energy of the homogeneous electron gas.
Surfaces are approximated by periodically repeated slabs. The InAs(111) 
surface, for example, is described by a supercell with a (2$\times$2) 
surface unit-cell and 10 atomic layers, the topmost 4 of them being 
fully relaxed. The atoms in the remaining layers are kept fixed at 
their bulk positions, using the theoretical bulk lattice constant of
5.98 \AA.
The bottom surfaces of the slabs are saturated with
hydrogen-like potentials.\cite{shiraishi:90} For the Ga-terminated surfaces 
a Coulomb potential with atomic number $Z=1.25$, screened by $Z$ electrons,
has been used, while a Coulomb potential with $Z=0.75$ has been taken for the
As-terminated surfaces. The saturated surfaces are semiconducting without
any surface states in the bulk band gap, in this way the interaction 
between both surfaces is made minimal.\cite{moll:96}
For the polar (111) surface of GaAs the uncertainty of the total 
energy due to charge transfer from one side of the slab to the opposite 
side has been estimated to be less than 1.4 meV/\AA$^2$.\cite{moll:96}
The In and As atoms are described by norm-conserving 
{\it ab initio} pseudopotentials\cite{bachelet:82}, which are further
transformed into fully separable Kleinman-Bylander\cite{kleinman:82}
pseudopotentials, with the $d$ potential chosen as the local potential. 
The wavefunctions are expanded into plane waves with a kinetic energy
$\le$ 10 Ry, and the electron density is calculated using special
${\bf k}$-point sets with the density in reciprocal space being equivalent 
to 64 ${\bf k}$-points in the whole (100)(1$\times$1) surface Brillouin 
zone. A generalized version of the computer code {\tt fhi93cp}\cite{stumpf:94}
has been employed. 

To derive a surface energy from a total-energy calculation 
both the top and the bottom surface of the slab have to be equivalent.
Such slabs can be constructed for, e.g., the (110) and the (100) orientation.
The (111) and (\=1\=1\=1) surfaces of InAs, however, are necessarily
inequivalent. This is an immediate consequence of the geometry
of the zinc-blende structure: the bulk lattice can be regarded as
a stacking of In--As double layers, which are cation and anion terminated 
towards the top and bottom surface of the slab, respectively.
Chetty and Martin\cite{chetty:92a} solved this problem by introducing an
energy density. While this energy density itself does not bear any
physical significance, they showed that integrals over suitably chosen subvolumes of
the supercell (e.g., volumes bounded by bulk mirror planes)
lead to well-defined, physically meaningful quantities. In this way
the total energy can be divided into a contribution from the upper
and the lower part of the slab, and after subtracting the respective volume
terms we get the surface energies of the top and bottom surface 
separately. This formalism has been implemented into the plane-wave
code.\cite{moll:96}

Similar surface energy calculations for GaAs are described in 
Ref.\,\onlinecite{moll:96}.  As the InAs surface reconstructions 
are expected to be equivalent to those found for GaAs, 
we have chosen the same candidates for low-energy surface structures as
discussed in that reference also for the present calculations. 
Furthermore, as epitaxial growth most often takes place under As-rich 
conditions, the energies in Tab. \ref{surface_energies} refer to 
As-rich conditions, i.e., surfaces which are in equilibrium with bulk As.

\begin{table}[bht]
\vspace{0.2cm}
\begin{center}
\begin{tabular}{|c|c|c|}
\hline
orientation & reconstruction                      & surface energy [meV/\AA$^2$] \\
\hline
(110)       & (1$\times$1) relaxed cleavage plane & 41  \\
(100)       & $\alpha$(2$\times$4)                & 48  \\
(100)       & c(4$\times$4)                       & 47  \\
(111)       & (2$\times$2) In vacancy             & 42  \\
(\=1\=1\=1) & (2$\times$2) As trimer              & 36  \\
\hline
\end{tabular}
\end{center}
\caption{
        The equilibrium surface reconstructions 
        of InAs under As-rich conditions and their surface energies.
        \label{surface_energies}
        }
\end{table}

As opposed to the GaAs\,(110) the As-terminated InAs\,(110) (1$\times$1) surface
does not become stable. Independent of the chemical 
environment, for InAs always the relaxed cleavage surface is energetically preferred, 
displaying the well-known outward 
rotation of the As atom (see e.g. the low-energy electron-diffraction analysis for 
the (1$\times$1) surface performed by Duke {\it et al.}\cite{duke:83}).

For the (100) orientation our calculation yields the c(4$\times$4) reconstruction 
as the lowest energy surface-structure under As-rich conditions. 
However, as can be read from 
Tab.\,\ref{surface_energies}, the energy difference with respect to the
$\alpha$(2$\times$4) reconstruction is so small that our calculation is
also compatible to the observation of a (2$\times$4) reconstruction. 
Experimentally\cite{yamaguchi:95} the surface reconstruction has been reported 
to change from (2$\times$4) to (4$\times$2) as a function of 
As chemical potential; in our computations the (4$\times$2)
reconstruction comes out somewhat too high in energy to give such a transition.
Further investigations are in progress.

Both for the (111) and the (\=1\=1\=1) orientation our predicted equilibrium 
reconstructions are consistent with recent core-level and valence 
photoemission studies.\cite{olsson:96a,andersson:96}
For the (111) the In-vacancy reconstruction is stable independently of the As
chemical potential. The As-trimer reconstruction, which in case of the GaAs(111) 
becomes the equilibrium structure under As-rich conditions, 
is too high in energy for InAs to be competitive and thus does not 
become stable.
On the (\=1\=1\=1) surface again both GaAs and InAs display the same 
As-trimer reconstruction in As-rich environment.

\begin{figure}[hbt]
\centerline{\psfig{figure=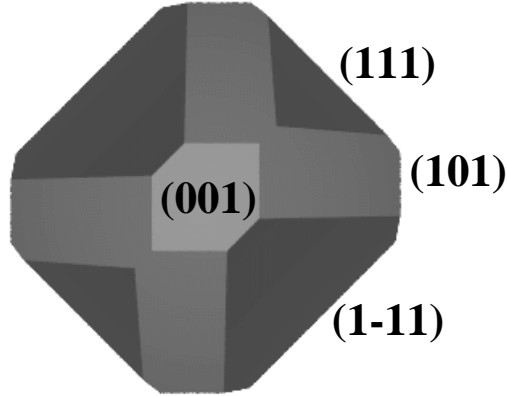,height=5.5cm}}
\caption{
        Equilibrium shape of InAs in an As-rich environment.
        Surfaces are labelled by their Miller indices.
        \label{ECS}
        }
\end{figure}

The equilibrium crystal shape (ECS) of InAs in As-rich environment 
displayed in Fig.\,\ref{ECS} has been derived by applying the Wulff 
construction to the data in Tab.\,\ref{surface_energies}.
As we know the surface energies only for the 
\{110\}, \{100\}, \{111\}, and \{\=1\=1\=1\} orientations we cannot exclude
that additional high Miller-index surface 
orientations may appear on the ECS, but the low Miller-index 
surfaces are expected to be the most prominent ones.
As a result we find that all four surface orientations co-exist on the ECS,
which means that they are thermodynamically stable with respect to
faceting into each other. This is in agreement with
the shape of large, and thus presumably fully relaxed, InAs islands 
grown on a GaAs substrate by metal-organic vapour-phase epitaxy
as observed by Steimetz {\it et al.}\cite{steimetz:96a}

\section*{\normalsize \bf 3. Elastic Relaxation Energy}

The equilibrium shape of strained three-dimensional islands 
grown on a lattice-mismatched substrate has to be carefully 
distinguished from the
equilibrium crystal shape described in the previous section. 
The optimum shape of a strained island, i.e., the shape that
corresponds to its lowest total energy, is additionally 
controlled by elastic relaxation.

We have computed the elastic energy within a continuum
theory. The strain field both in the island and in the bulk is
fully accounted for. For simplicity we have taken the experimental 
second and third order elastic moduli for GaAs\cite{data:91} to describe 
the elastic properties of both the 
island and the substrate. A test calculation with the linear 
elastic constants of InAs and GaAs has shown that this approximation 
does not affect any of our qualitative conclusions drawn with 
respect to the equilibrium shape. A finite element approach is applied 
to compute the displacement field ${\bf u}({\bf r})$ and the
strain tensor $\epsilon({\bf r})$: The island and the slab
representing the semiinfinite substrate are divided into small 
irregular tetrahedra.
The displacement field is defined on the vertices of this
subdivision, and in between a piecewise linear interpolation
of ${\bf u}({\bf r})$ is used. Thus the strain field $\epsilon({\bf r})$
calculated from ${\bf u}({\bf r})$ is constant within each of the
small tetrahedra. The elastic energy, which is calculated by summing the 
elastic energy density $f(\epsilon)$ times the volume over all
tetrahedra, is iteratively minimized with respect to ${\bf u}({\bf r})$.
Above procedure is repeated for several finenesses of the subdivision,
and the energies are extrapolated to fineness equal to zero.

The elastic energy $E_{\rm{mesa}}$ of a truncated pyramid
with volume $V$ can be approximated by
a simple analytic expression once the elastic energy of the 
pyramid with the same side faces is known.
Let $E_{\rm{pyr}}$ denote the elastic energy of this pyramid with volume $V$.
>From the variational property of the elastic energy with respect to the
displacement field $u({\bf r})$ 
the elastic energy of the truncated pyramid $E'$ of volume $V'$ is estimated by
\begin{equation}
E' \le E_{\rm{pyr}} - \int_{V \setminus V'} f(\epsilon({\bf r}))d^3{\bf r},
\end{equation}
with the integral denoting the elastic energy of that part of the pyramid which
has been sliced off.
In the following we are even going to neglect this integral. This still
yields a good approximation to $E'$ for any not too flat object, 
because the tops of the islands are almost fully relaxed.
Finally, we make use of the scaling property of the elastic energy, 
$E(V) \sim V$, to transform the energy from $V'$ back to the island volume $V$:
\begin{equation}
E_{\rm{mesa}} \le V / V'\, ( E_{\rm{pyr}} - 
  \int_{V \setminus V'} f(\epsilon({\bf r}))d^3{\bf r} ).
\end{equation}
The quality of this approximation can be judged from Fig.\,\ref{optimization}.
The full lines, given by $V / V' E_{\rm{pyr}}$, pass through the 
data points (diamonds and squares, respectively) computed for truncated 
pyramids with two \{111\} and two \{\=1\=1\=1\} or four \{101\} faces.

\section*{\normalsize \bf 4. Equilibrium Island Shape}

To illustrate the basic physical mechanism we are first going to 
restrict ourselves to a very small part of configuration space, i.e.,
we consider a square based pyramid with four \{101\} facets
and compare to the related truncated pyramids which are generated 
by slicing off the top, thus creating a mesa-shaped island with a (001) 
plane on top, and rescaling the lengths such that the volume is kept 
equal to the volume of the original pyramid.

\begin{figure}[htb]
\centerline{\psfig{figure=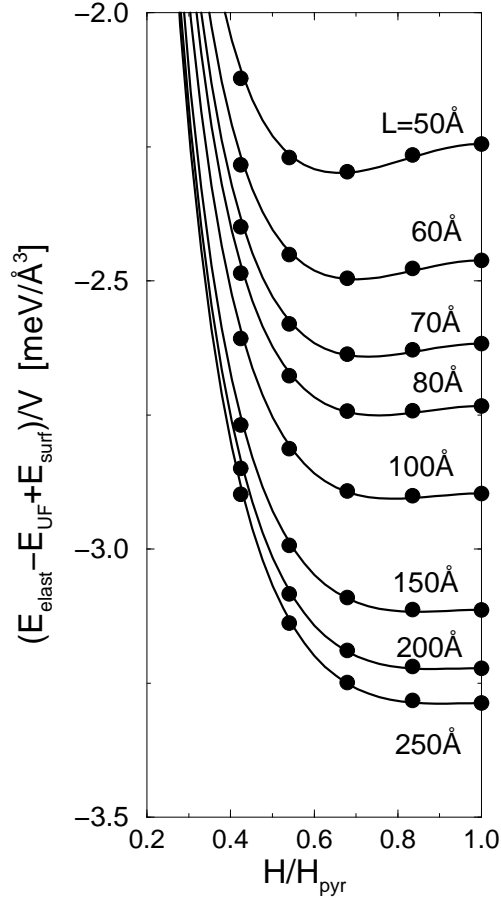,height=12.0cm}}
\caption{
        Energy gain due to formation of coherent islands (i.e., 
        elastic energy $E_{\rm{elast}}$ 
        plus surface energy $E_{\rm{surf}}$ 
        minus energy of the uniform strained film $E_{\rm{UF}}$)
        vs. height $H$ of the truncated pyramid, normalized 
        to the height $H_{\rm{pyr}}$ of the pyramid with the same volume.
        The truncated pyramids are bounded by four \{110\} 
	facets and one (001) facet. 
        The energy curves are parameterized by 
        the width $L$ of the pyramid, which is just a simple measure 
        for the volume.
        The filled circles denote computed values, while the lines represent
        the analytic approximation discussed in the previous section.
        \label{qdot_101_total_energies}
        }
\end{figure}

When a truncated pyramid is created from a pyramid in this way the elastic energy
has to increase, because material is taken away from the top of the island,
where it is already almost fully relaxed, and deposited in the remaining mesa
where it is still considerably strained.
On the other hand, given the InAs-ECS from Sect.\ 2, the surface energy
decreases. The optimum configuration therefore results from a 
competition between the elastic relaxation energy and the surface energy.
The scaling properties of these two quantities, however, are
different, the elastic energy being proportional to the volume while the
surface energy increases with volume like $V^{2/3}$. Therefore 
the surface energy gains more importance at small volume, while the elastic 
energy dominates at large volume.\footnote{From
a practical point of view the volume interval where scaling makes sense
is of course limited by atomistic effects for $V \rightarrow 0$ and
the generation of dislocations at large volume $V$.}
This explains the results displayed in Fig.\,\ref{qdot_101_total_energies}:
For small volume the total energy as a function of the height of the
truncated pyramid has a minimum corresponding to some mesa-type shape.
When the volume of the island increases, the elastic relaxation energy 
gains more influence and therefore this minimum becomes 
less pronounced and shifts towards the pyramidal geometry.

\begin{figure}[htb]
\centerline{\psfig{figure=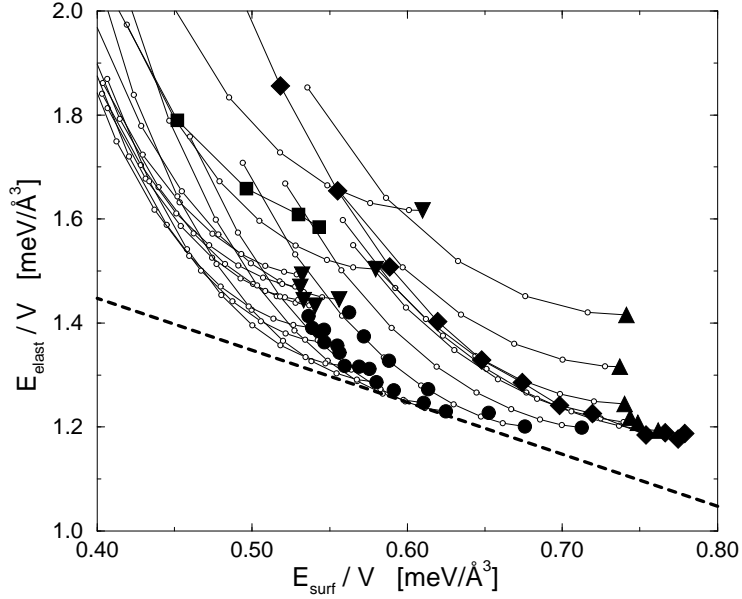,height=8.0cm}}
\caption{Elastic energy per volume $E_{\rm{elast}}/V$  vs. surface energy per volume
         $E_{\rm{surf}}/V$ for InAs islands with volume $V=2.88\times 10^5$\AA$^3$.
         Squares: square based pyramid with four \{101\} faces and (001)-truncated 
	 \{101\}-pyramids.
         Diamonds: square based pyramids with two \{111\} and two \{\=1\=1\=1\} 
         faces and (001)-truncated pyramids.
         Triangles up: ``huts'' with two \{111\} and two \{\=1\=1\=1\} faces.
         Triangles down: square based \{101\} pyramids with \{\=1\=1\=1\}-truncated edges.
         Dots: islands with four \{101\}, two \{111\}, and two \{\=1\=1\=1\} faces.
         Filled symbols denote numerical results, while open circles correspond to a
         simple analytical approximation for (001)-truncated ``mesa-shaped'' islands,
         assuming that the elastic energy does not change when the
         (almost fully relaxed) top of an island is cut off.
         Full lines connect islands that are created in this
	 way, varying the height of the
         (001) surface plane.
         The dashed line is the curve of constant total energy $E_{\rm{elast}} + E_{\rm{surf}}$
         that selects the equilibrium shape.
         \label{optimization}
        }
\end{figure}

To derive the equilibrium island shape we have to account for the whole variety
of possible island configurations. To this purpose we have calculated 
the elastic and surface energies for various arbitrarily shaped 
InAs islands bounded by \{101\}, \{111\}, and \{\=1\=1\=1\} facets. 
The results are displayed by filled symbols in Fig.\,\ref{optimization}. 
The elastic energies of the related (001)-truncated ``mesa-shaped'' islands have been 
derived by means of Eq.\,(2) as described in Sect. 3, the respective data being 
denoted by open circles in Fig.\,\ref{optimization}.
The optimum shape at given volume corresponds to that
point where the line of constant total energy touches the manifold
of island energies from below.
Even when the volume is changed, Fig.\,\ref{optimization} can nevertheless still be used
to derive the optimum shape: From the scaling relations we know
that the ordinate does not change, while the abscissa has to be rescaled
according to $E_{\rm{surf}}/V \sim V^{-1/3}$, i.e., only the slope of the 
total energy line decreases when the volume increases. 

\begin{figure}[htb]
\centerline{\psfig{figure=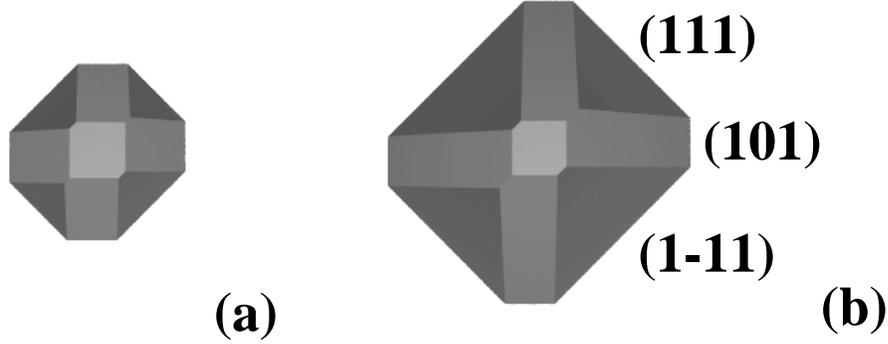,height=5.5cm}}
\caption{
        The equilibrium shape of a strained coherent InAs island 
        for two different volumes, 
        (a) $V \sim 8 \times$ 10$^4$ \AA$^3$,
        (b) $V \sim 36 \times$ 10$^4$ \AA$^3$. 
        The orientation of the coordinate system is identical to Fig.\,\ref{ECS}. 
         \label{qdot_vgl}
        }
\end{figure}

The equilibrium island shapes (see Fig.\,\ref{qdot_vgl}) are mesa-like
hills bounded by \{101\}, \{111\}, and \{\=1\=1\=1\} facets and
a (001) surface on the top, similar to a truncated ECS. 
Due to the different scaling properties
of $E_{\rm{elast}}$ and $E_{\rm{surf}}$ the islands prefer a steeper
and more pyramidal shape at larger volume while they tend to be flatter at small
volume. E.g., in Fig.\,\ref{qdot_vgl} it can be seen that in comparison to
the steeper \{111\} faces the \{101\} faces are more extended
on the small island than on the larger one.  
Our equilibrium island shapes differ from the \{101\}-pyramids grown 
experimentally by Ruvimov {\it et al.}\cite{ruvimov:95} for yet unknown reason.
They are, however, similar to the shapes of InP islands on GaInP observed by
Georgsson {\it et al.}\cite{georgsson:95} 

The equilibrium island shape evolves continuously with respect to the volume. 
Therefore the total energy of the equilibrium islands does not simply 
vary with volume like $a + b V^{-1/3}$ as would follow from simple scaling.
Instead, when we represent the
low-energy envelope of the data in Fig.\,\ref{optimization} 
by $y = a + b x^{-\gamma}$, with $x = E_{\rm{surf}}/V^{2/3}$ and
$y = E_{\rm{elast}}/V$,  and $a$, $b$, and $\gamma$ 
independent of volume, $b$ and $\gamma$ positive, 
we get the total energy per volume from the Legendre-transform of $y(x)$:
\begin{equation}
E_{\rm{tot}} / V = a + b'\, V^{-\gamma/3(\gamma+1)},
\end{equation}
with some positive constant $b'$, i.e., the volume-exponent becomes smaller.
Of course our approach which, e.g., disregards surface
stress effects, still leads to Ostwald ripening by means of its construction.
To discuss the possibility of an optimum
island size one has to consider further elastic 
interactions.\cite{shchukin:95a} 

\section*{\normalsize {\bf Acknowledgment}}

We thank E. Steimetz for helpful discussion and a copy 
of Ref.\,\onlinecite{steimetz:96a} prior to
publication. This work was supported in part by the Sfb 296 of the 
Deutsche Forschungsgemeinschaft.

\bibliographystyle{prsty}
\bibliography{/home/pehlke/tex/bib/physics}

\begin{thebibliography}{10}

\bibitem{eaglesham:90}
D.~J. Eaglesham and M. Cerullo, Phys. Rev. Lett. {\bf 64},  1943  (1990).

\bibitem{vanderbilt:91}
D. Vanderbilt and L.~K. Wickham,  in {\em Evolution of Thin-Film and Surface
  Microstructure, MRS Symposia Proceedings No. 202}, edited by C.~V. Thompson,
  J.~Y. Tsao, and D.~J. Srolovitz (Material Research Society, Pittsburgh,
  1991), p.\ 555.

\bibitem{ledentsov:96}
N.~N. Ledentsov,  in {\em Proceedings of the 23rd International Conference on
  the Physics of Semiconductors}, edited by M. Scheffler and R. Zimmermann
  (World Scientific, Singapore, 1996), p.\ 19.

\bibitem{snyder:91}
C.~W. Snyder, B.~G. Orr, D. Kessler, and L.~M. Sander, Phys. Rev. Lett. {\bf
  66},  3032  (1991).

\bibitem{leonard:93}
D. Leonard {\it et~al.}, Appl. Phys. Lett. {\bf 63},  3203  (1993).

\bibitem{moison:94}
J.~M. Moison {\it et~al.}, Appl. Phys. Lett. {\bf 64},  196  (1994).

\bibitem{ruvimov:95}
S. Ruvimov {\it et~al.}, Phys. Rev. B {\bf 51},  14766  (1995).

\bibitem{bressler:95}
V. Bressler-Hill {\it et~al.}, Phys. Rev. Lett. {\bf 74},  3209  (1995).

\bibitem{petroff:96}
P.~M. Petroff and G. Medeiros-Ribeiro, MRS Bulletin {\bf 21},  50  (1996).

\bibitem{grundmann:95a}
M. Grundmann {\it et~al.}, Phys. Rev. Lett. {\bf 74},  4043  (1995).

\bibitem{grundmann:95b}
M. Grundmann, O. Stier, and D. Bimberg, Phys. Rev. B {\bf 52},  11969  (1995).

\bibitem{kirstaedter:94}
N. Kirstaedter {\it et~al.}, Electronics Letters {\bf 30},  1416  (1994).

\bibitem{priester:95}
C. Priester and M. Lannoo, Phys. Rev. Lett. {\bf 75},  93  (1995).

\bibitem{shchukin:95a}
V.~A. Shchukin, N.~N. Ledentsov, P.~S. Kop'ev, and D. Bimberg, Phys. Rev. Lett.
  {\bf 75},  2968  (1995).

\bibitem{jesson:96a}
D.~E. Jesson, K.~M. Chen, and S.~J. Pennycook, MRS Bulletin {\bf 21},  31
  (1996).

\bibitem{perdew:81}
J.~P. Perdew and A. Zunger, Phys. Rev. B {\bf 23},  5048  (1981).

\bibitem{ceperley:80}
D.~M. Ceperley and B.~J. Alder, Phys. Rev. Lett. {\bf 45},  566  (1980).

\bibitem{shiraishi:90}
K. Shiraishi, J. Phys. Soc. Jap. {\bf 59},  3455  (1990).

\bibitem{moll:96}
N. Moll, A. Kley, E. Pehlke, and M. Scheffler, Phys. Rev. B {\bf 54},  8844
  (1996).

\bibitem{bachelet:82}
G.~B. Bachelet, D.~R. Hamann, and M. Schl\"uter, Phys. Rev. B {\bf 26},  4199
  (1982).

\bibitem{kleinman:82}
L. Kleinman and D.~M. Bylander, Phys. Rev. Lett. {\bf 48},  1425  (1982).

\bibitem{stumpf:94}
R. Stumpf and M. Scheffler, Comput. Phys. Commun. {\bf 79},  447  (1994).

\bibitem{chetty:92a}
N. Chetty and R.~M. Martin, Phys. Rev. B {\bf 45},  6074  (1992).

\bibitem{duke:83}
C.~B. Duke, A. Paton, A. Kahn, and C.~R. Bonapace, Phys. Rev. B {\bf 27},  6189
   (1983).

\bibitem{yamaguchi:95}
H. Yamaguchi and Y. Horikoshi, Phys. Rev. B {\bf 51},  9836  (1995).

\bibitem{olsson:96a}
L.~O. Olsson {\it et~al.}, Phys. Rev. B {\bf 53},  4734  (1996).

\bibitem{andersson:96}
C.~B.~M. Andersson {\it et~al.}, Surf. Sci. {\bf 347},  199  (1996).

\bibitem{steimetz:96a}
E. Steimetz, F. Schienle, J.-T. Zettler, and W. Richter, J. Cryst. Growth (in press).

\bibitem{data:91}
{\em Data in Science and Technology, Semiconductors, Group IV Elements and
  III-V Compounds}, edited by O. Madelung (Springer, Berlin, 1991).

\bibitem{georgsson:95}
K. Georgsson {\it et~al.}, Appl. Phys. Lett. {\bf 67},  2981  (1995).

\end{thebibliography}
\end{document}